\documentclass[12pt,preprint]{aastex}

\def\ltsima{$\; \buildrel < \over \sim \;$}
\def\gtsima{$\; \buildrel > \over \sim \;$}
\def\lsim{\lower.5ex\hbox{\ltsima}}
\def\gsim{\lower.5ex\hbox{\gtsima}}
\def\lapp{\ifmmode\stackrel{<}{_{\sim}}\else$\stackrel{<}{_{\sim}}$\fi}
\def\gapp{\ifmmode\stackrel{>}{_{\sim}}\else$\stackrel{<}{_{\sim}}$\fi}

\newdimen\minuswidth    
\setbox0=\hbox{$-$}
\minuswidth=\wd0
\catcode`@=\active
\def@{\kern\minuswidth}
\setbox0=\hbox{\rm0}
 
\shorttitle{Blue Stragglers in M55}
\shortauthors{Lanzoni et al.}
 
\begin{document} 
 
\title{The surprising external upturn of the Blue Straggler radial
distribution in M55\altaffilmark{1}}

\author{
B. Lanzoni\altaffilmark{2},
E. Dalessandro\altaffilmark{2,3},
S. Perina\altaffilmark{2,4},
F.R. Ferraro\altaffilmark{2},
R.T. Rood\altaffilmark{5},
A. Sollima\altaffilmark{2}
}
\altaffiltext{1}{Based on observations with the NASA/ESA {\it HST}, obtained at the
Space Telescope Science Institute, which is operated by AURA, Inc., under
NASA contract NAS5-26555. Also based on WFI observations collected
at the European Southern Observatory, La Silla, Chile.}
\altaffiltext{2}{Dipartimento di Astronomia, Universit\`a degli Studi
di Bologna, I--40127 Bologna, Italy}
\altaffiltext{3}{Agenzia Spaziale Italiana, Centro di Geodesia
Spaziale, I--75100 Matera, Italy}
\altaffiltext{4}{INAF--Osservatorio Astronomico di Bologna, I--40127
Bologna, Italy}
\altaffiltext{5}{Astronomy Department, University of Virginia,
P.O. Box 400325, Charlottesville, VA, 22904}
\date{ApJ accepted, 8 August, 07}

\begin{abstract}
By combining high-resolution {\it HST} and wide-field ground based
observations, in ultraviolet and optical bands, we study the Blue Straggler
Star (BSS) population of the low density galactic globular cluster M55
(NGC~6809) over its entire radial extent.  The BSS projected radial distribution is
found to be bimodal, with a central peak, a broad minimum at intermediate
radii, and an upturn at large radii. Similar bimodal distributions have been
found in other globular clusters (M3, 47~Tucanae, NGC~6752, M5), but the
external upturn in M55 is the largest found to date.  This might indicate a
large fraction of primordial binaries in the outer regions of M55, which
seems somehow in contrast with the relatively low ($\sim 10\%$) binary
fraction recently measured in the core of this cluster.
\end{abstract}


\section{INTRODUCTION}
Blue straggler stars (BSS) are objects that the single mass stellar evolution
theory is unable to explain. In the color-magnitude diagrams (CMDs) of
evolved stellar populations, like globular clusters (GCs), they lie along an
extension of the Main Sequence (MS), in a region which is brighter and bluer
than the Turn-Off (TO), where no stars are expected to be found.  Their
position in the CMD indicates that BSS are rejuvenated stars, with masses
larger than the normal cluster stars \citep[this is also confirmed by direct
mass measurements; e.g.][]{sha97}.  Thus, they are thought to have increased
their initial mass during their evolution, and two main scenarios have been
proposed for their formation: the {\it collisional scenario} \citep{colbss}
suggests that BSS are the end-products of stellar mergers induced by
collisions (COL-BSS), while in the \emph{mass-transfer} \citep{mtbss1,
mtbss2} scenario BSS form by the mass-transfer activity between two
companions in a binary system (MT-BSS), possibly up to the complete
coalescence of the two stars.  Hence, understanding the origin of BSS in
stellar clusters provides valuable insight both on the binary evolution
processes, and on the effects of dynamical interactions on the (otherwise
normal) stellar evolution.

The two formation mechanisms are likely to be at work
simultaneously in every GC \citep[see the case of M3 as an
example;][1997]{fe93}, with a relative efficiency that probably
depends on the local density \citep{fp92, fe99a, bel02, fe03}. In
fact, since stellar collisions are most probable in high-density
environments, COL-BSS are expected to be formed preferentially in
the cluster cores, while MT-BSS can populate both the centre and the
peripheries. Primordial binaries can in fact sink to the core due to
dynamical friction and mass segregation processes, and ``new''
binaries can be formed in the cluster centers by gravitational
encounters. To be completely clear with our terminology, a
primordial binary which has sunk to the cluster center and is then
driven to merge by stellar interactions (rather than by evolution of
the more massive member of the binary or by magnetic braking as in
the case of W Uma systems)  is classified as a COL-BSS. Of course,
in the case of a low-density cluster a fraction of primordial
binaries evolving in isolation  (hence classified as MT-BSS in our
terminology) can well be present even in the cluster center.

One possibility for distinguishing between the two types of BSS is offered by
high-resolution spectroscopic studies. In fact, anomalous chemical abundances
are expected at the surface of BSS resulting from MT activity
\citep{sarna96}, while they are not predicted in case of a collisional
formation \citep{lomb95}.  Such studies have just become feasible, and the
results found in the case of 47~Tucanae \citep[47~Tuc;][]{COdep} are
encouraging.
The detection of unexpected properties of stars along standard evolutionary
sequences (e.g., variability, anomalous population fractions, peculiar radial
distributions, or a secondary MS) can help to estimate the fraction of
binaries within a cluster \citep[see, e.g.,][]{bail94, albr01, bel02, bec06b,
sollima07}, but such evidence does not directly allow the determination of
the relative efficiency of the two BSS formation processes.
Instead, the observational study of the BSS radial distribution within the
host clusters, complemented with suitable dynamical simulations, has proved
to be a more widely applicable and powerful tool \citep[see][for a
review]{fe06}. Observations have shown that BSS are generally highly
concentrated in the cluster cores, and in some cases, specifically in M3
\citep{fe97}, 47~Tuc \citep{fe04}, NGC~6752 \citep{sab04}, and M5 \citep{w06,
lan07a}, their projected radial distribution is bimodal, i.e., their fraction with
respect to the normal cluster populations (like horizontal branch or red
giant branch stars) decreases to a minimum, and then rises again to larger
values for increasing radii.  Dynamical simulations suggest that the external
rising branch cannot be due to COL-BSS generated in the core and kicked out
by dynamical interactions \citep{ma04}. Instead, the bimodality of the radial
distribution can be explained \citep[Mapelli et al. 2004, 2006;][]{lan07a} by
assuming that a non-negligible fraction ($\gsim 20\%-40\%$) of the BSS
population is made of MT-BSS (responsible for the external rising branch),
with the balance being COL-BSS (mainly contributing to the central peak).
The atypical case of $\omega~$Centauri \citep[where the BSS radial
distribution has been found to be flat;][]{fe06b} can be explained if
the mass segregation processes have not yet played a major role in this
system, thus implying that it is populated by a vast majority of MT-BSS
\citep{ma06}.  These results demonstrate that detailed studies of the BSS
radial distribution within GCs are powerful tools for better understanding
the complex interplay between dynamics and stellar evolution in dense stellar
systems.  Extending this kind of investigation to a larger sample of GCs,
with different structural and dynamical characteristics, is crucial for
identifying the cluster properties that mainly affect the BSS formation
mechanisms and their relative efficiency.

The present paper is devoted to the study of the BSS projected radial distribution in
M55 (NGC~6809).  Previous works have suggested that BSS in this cluster are
more centrally concentrated than MS and sub-giant branch (SGB) stars
\citep[Mandushev et al. 1997;][hereafter Z97]{zag97}, and that the BSS radial
distribution is bimodal (Z97).  These studies, however, were based only on
partial coverage of the cluster area, while the wide-field observations used
in this paper cover almost all of the entire cluster extension. In addition,
we have sampled the core region with {\it HST} high-resolution observations
both in the ultraviolet and in the optical bands, thus getting a more
reliable and efficient detection of the BSS and of the normal cluster
populations.  In the following we present the observational data sets and the
results obtained. We postpone to a forthcoming paper the theoretical
interpretation of the BSS radial distribution by means of dynamical
simulations, and a detailed comparison with all the other GCs studied with
the same technique.

\section{OBSERVATIONS AND DATA ANALYSIS}
\subsection{The Data Sets}
\begin{figure}[!hp]
\begin{center}
\includegraphics[scale=0.7]{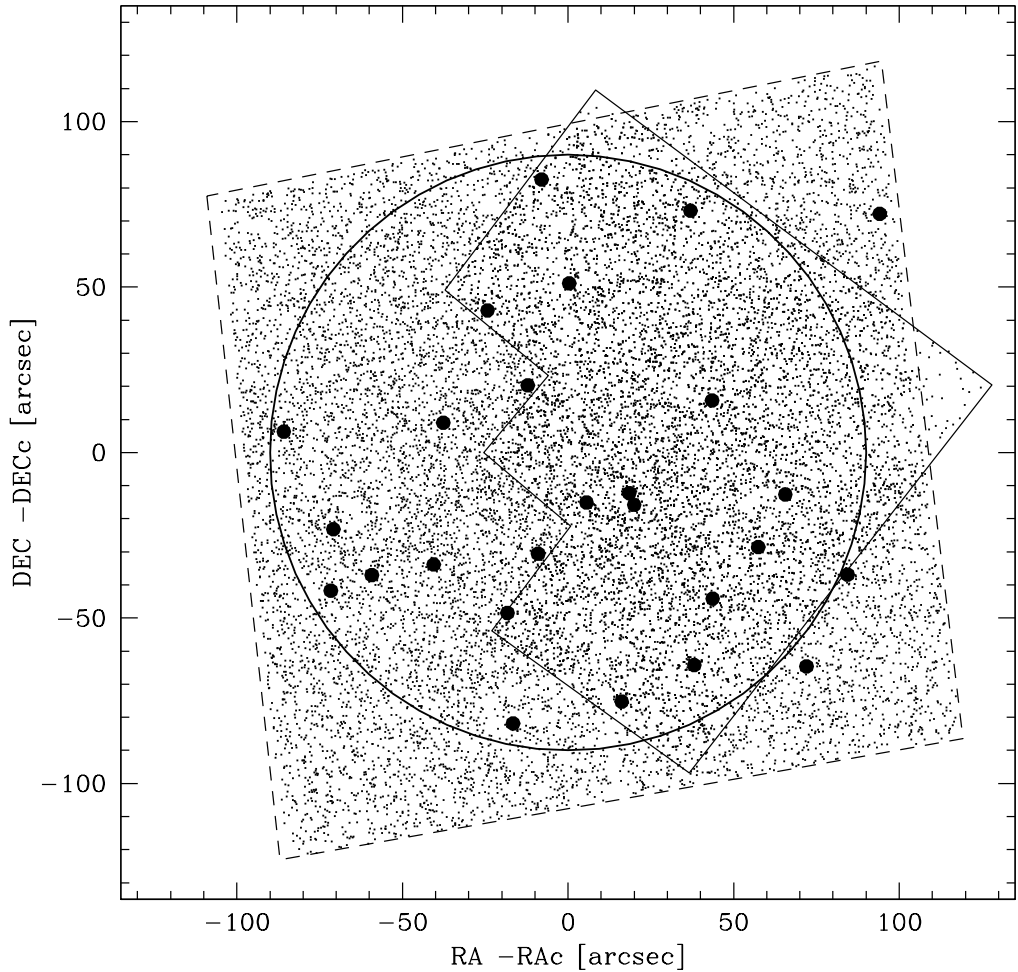}
\caption{Map of the {\it HST} sample, with coordinates referred to the
derived center of gravity $C_{\rm grav}$: RA$_{\rm c}= 19^{\rm h}\, 39^{\rm
m}\, 59\fs 54$, DEC$_{\rm c}= -30^{\rm o}\, 57\arcmin\, 45\farcs 14$. The
solid and dashed thin lines delimit the {\it HST}-WFPC2 and {\it HST}-ACS
fields of view, respectively.  The selected BSS (heavy dots) and the annulus
with radius $r=90\arcsec$ used to study their projected radial distribution (cfr. Table
\ref{tab:annuli}) are also shown.}
\label{fig:HST}
\end{center}
\end{figure}

The present study is based on a combination of two different photometric data
sets:

\emph{1. The high-resolution set} -- It consists of a series of {\it
HST}-WFPC2 images of the cluster center (Prop. 10524, P.I. Ferraro), obtained
through filter F255W (medium UV, for a total exposure time $t_{exp}=2000$ s)
and F336W (approximately corresponding to an $U$ filter, with $t_{exp}=1600$
s).  To efficiently resolve the stars in the highly crowded central regions,
the Planetary Camera has been pointed approximately on the cluster
center, while the three Wide Field Cameras (WFC) have been used to sample the
surrounding regions.  The photometric reduction of the images was carried out
using ROMAFOT \citep{buon83}, a package developed to perform accurate
photometry in crowded fields and specifically optimized to handle
under-sampled point spread functions \citep[PSFs;][]{buon89}, as in the case
of the WFC chips.
Additional {\it HST} images of the cluster center, obtained with the ACS-Wide
Field Channel (Prop. 10775, P.I. Sarajedini) have been retrieved from the
ESO-STECF Science Archive. Only the short exposures (10 sec each) in filters
F606W ($V$) and F814W ($I$) have been used in the present work. The adopted
data reduction procedure is described in detail in \citet{sollima07}.  The
map of the ACS data set is shown in Figure \ref{fig:HST}, together with the
WFPC2 and ACS fields of view (FoVs).

\emph{2. The wide-field set} - A complementary set of public wide-field $B$
and $V$ images obtained with the Wide Field Imager (WFI) at the 2.2m ESO-MPI
telescope was retrieved from the ESO Science Archive.  Thanks to the wide
($34\arcmin\times 34\arcmin$) FoV of WFI, 
these data almost cover the entire cluster extension (see Fig.
\ref{fig:WFI}, where the cluster is roughly centered on CCD $\# 7$).
The raw WFI images were corrected for bias and flat field, and the overscan
regions were trimmed using IRAF\footnote{IRAF is distributed by the National
Optical Astronomy Observatory, which is operated by the Association of
Universities for Research in Astronomy, Inc., under cooperative agreement
with the National Science Foundation.} tools. The PSF fitting procedure was
then performed independently on each image using DoPhot \citep{dophot}.

\subsection{Astrometry and Photometric Calibration}
The {\it HST} and WFI catalogs have been placed on the absolute astrometric
system by adopting the procedure described in \citet[][2003]{fe01}.  The new
astrometric Guide Star Catalog (GSC-II\footnote{Available at {\tt
http://www-gsss.stsci.edu/Catalogs/GSC/GSC2/GSC2.htm}.}) was used to search
for astrometric standard stars in the WFI FoV, and a specific
cross-correlation tool has been employed to obtain an astrometric solution
for each of the 8 CCDs.  Several hundred GSC-II reference stars were found in
each chip, thus allowing an accurate absolute positioning.  Then, a few
hundred stars in common between the WFI and the {\it HST} FoVs have been used
as secondary standards to place the {\it HST} catalogs on the same absolute
astrometric system.  At the end of the procedure the global uncertainties in
the astrometric solution are of the order of $\sim 0\farcs 2$, both in right
ascension ($\alpha$) and declination ($\delta$).

The photometric calibration of the optical ($B$ and $V$) magnitudes has been
performed by using the Stetson Photometric Standard
catalog\footnote{Available at {\tt
http://www1.cadc-ccda.hia-iha.nrc-cnrc.gc.ca/community/STETSON/standards/}.}.
After cross correlating the WFI and Stetson catalogs, we have used the stars
in common for the calibration of the WFI $B$ and $V$ magnitudes.  Then, the
ACS $V$ magnitudes have been converted to the WFI system by using the stars
in common. Since the Stetson standard field does not overlap with the ACS
FoV, the calibration of the ACS $I$ magnitudes has been performed by using
the stars in common with the catalog of \citet{desidera}, after converting
the latter to the Stetson photometric system. Finally, the WFPC2 $m_{255}$
and $U$ magnitudes have been calibrated to the \citet{holtz95} zero-points.
The resulting CMDs, both in the UV and optical bands, are shown in Figures
\ref{fig:uvCMD} and \ref{fig:optCMD}, respectively.

Unless otherwise specified, in the following analysis we adopt the combined
{\it HST} catalog (ACS and WFPC2 data) for the cluster central regions (see
Fig. \ref{fig:HST}), and the complementary WFI sample for the external parts
(see Fig. \ref{fig:WFI}).

\subsection{Center of Gravity, and Density Profile}
\label{sec:prof}
Given the absolute positions and the magnitudes of individual stars, the
center of gravity $C_{\rm grav}$ has been determined by averaging the
coordinates $\alpha$ and $\delta$ of all stars brighter than $V = 19$ lying
in the FoV of WFI CCD $\# 7$.  We have chosen to use the WFI (instead of the
{\it HST}) data, because in such a loose cluster the FoV of the WFPC2
planetary camera is too small to provide an adequately large sample for the
accurate determination of the center of gravity, while the ACS FoV is crossed
by the gap between the two chips.  Following the iterative procedure
described in \citet{mont95}, we have determined $C_{\rm grav}$ to be
$\alpha({\rm J2000}) = 19^{\rm h}\, 39^{\rm m}\, 59\fs 54$, $\delta ({\rm
J2000})= -30^{\rm o}\, 57\arcmin\, 45\farcs 14$, with a 1$\sigma$ uncertainty
of $0\farcs 5$ in both $\alpha$ and $\delta$.  This value of $C_{\rm grav}$
is located $\sim 2\arcsec$ south-east ($\Delta\alpha = 2\farcs 1$,
$\Delta\delta=-1\farcs 1$) from that previously derived by \citet{har96} on
the basis of the surface brightness distribution.

By exploiting the optimal combination of high-resolution and wide-field
sampling provided by our observations, we have determined the projected
density profile by direct star counts over the entire cluster radial extent,
from $C_{\rm grav}$ out to $\sim 1400\arcsec\sim23\arcmin$.  To avoid biases
due to incompleteness, we have considered only stars brighter than $V= 19$
from the ACS and the complementary WFI catalogs (see Fig. \ref{fig:optCMD}).
The brightest red giant branch (RGB) stars that are strongly saturated in the
ACS data set have been excluded from the analysis, but since they are few in
number and the ACS pixel scale is only of $0.05\arcsec/pixel$, the effect on
the resulting density profile is negligible.  Following the procedure
described in \citet[][2004]{fe99a}, we have divided the entire {\it HST}+WFI
sample in 26 concentric annuli, each centered on $C_{\rm grav}$ and split in
an adequate number of sub-sectors. The number of stars lying within each
sub-sector was counted, and the star density was obtained by dividing these
values by the corresponding sub-sector areas.  The stellar density in each
annulus was then obtained as the average of the sub-sector densities, and its
standard deviation was estimated from the variance among the sub-sectors.
The radial density profile thus derived is shown in Figure \ref{fig:prof},
and the average of the three outermost ($r\gsim 17\arcmin$) measures has been
used to estimate the background contribution (corresponding to $\sim 3$ stars
arcmin$^{-2}$). Figure \ref{fig:prof} also shows the best-fit mono-mass King
model and the corresponding values of the core radius and concentration:
$r_c=114\arcsec$ (with a typical error of $\sim \pm 2\arcsec$) and $c=1$,
respectively.  These values are in agreement with those quoted by
\citet[][$r_c=126\farcs 4$ and $c=0.93$]{mcL05}, and by \citet[][$r_c\sim
120\arcsec$ and $c\sim 1$]{irwtrim84}. Concentration parameters as low as
$\sim 0.8$ \citep[as quoted, e.g., by][Z97] {har96} provide significantly
worse fits to the observed profile. The difference with respect to (Z97, who
also computed the surface density profile by direct star counts) is probably
due to the fact that their ground-based observations are saturated at $V\lsim
14$, and have a pixel scale much larger than that of ACS, so they have lost a
number of faint stars in the central regions of the cluster.  Assuming the
distance modulus $(m-M)_0=13.82$ \citep[$d\sim 5.8$ Kpc,][]{fe99b}, our value
of $r_c$ corresponds to $\sim 3.2$ pc. These values can then be used to
redetermine the other structure parameters of the cluster.  By assuming
$\mu_0=19.13$ mag/arcsec$^2$ as the central surface brightness \citep{har96},
and $E(B-V)=0.07$ as reddening \citep{fe99b}, we estimate that the
extinction-corrected central surface brightness of the cluster is
$\mu_{V,0}(0)\simeq 18.91$ mag/arcsec$^2$.  Following the procedure described
in \citet[][see also Beccari et al. 2006a]{djorg93}, we derive $\log\nu_0
\simeq 2.23$, where $\nu_0$ is the central luminosity density in units of
$L_\odot\,{\rm pc}^{-3}$. By assuming a mass-to-light ratio $M/L_V=3$, the
derived central mass density measured in $M_\odot/$pc$^3$ is
$\log\rho_0=2.7$, which is a factor $\sim 1.6$ higher than that quoted by
\citet{pry93}. This value corresponds to $n_0\simeq 1000$ stars pc$^{-3}$ if
a mean stellar mass of $0.5\,M_\odot$ is assumed.

\section{CLUSTER POPULATION SELECTION}
\label{sec:samples}

\subsection{The BSS Population}
At UV wavelengths BSS are among the brightest objects in a GC, and RGB stars
are particularly faint.  By combining these advantages with the
high-resolution capability of {\it HST}, the usual problems associated with
photometric blends and stellar crowding
are minimized, and BSS can be most reliably recognized and separated from the
other populations in the UV CMDs \citep[see also][]{fe04}.
For these reasons our primary criterion for the definition of the BSS sample
is based on the position of stars in the ($m_{255},~m_{255}-U$) plane.  In
order to avoid incompleteness bias and to limit the possible contamination
from TO and SGB stars, we have chosen a limiting magnitude of $m_{255}=18.5$
(roughly 1 magnitude brighter than the cluster TO). The adopted selection box
and the resulting 12 BSS identified in the UV plane are shown in Figure
\ref{fig:uvCMD}.
Once selected in the UV CMD, the BSS lying in the field in common with the
ACS sample have been used to define the selection box and the limiting
magnitude in the ($V,~V-I$) plane.  The latter is $V\simeq
17.5$, and the adopted selection box is shown in the left-hand panel of
Figure \ref{fig:optCMD}. One of the BSS candidates (that lies close to the
reddest edge of the box) has been rejected from the sample on the basis of
its position in the UV plane, where it is $\sim 0.2$ magnitudes fainter than
the adopted $m_{255}$ limit and has a color of $m_{255}-U = 1$, thus clearly
belonging to the SGB star population. A total of 24 BSS have been identified
within the ACS selection box, of which 11 are in common with the WFPC2
sample.
Finally, in order to select the BSS population in the complementary
WFI data set, we have adopted the same $V$ magnitude limits as for the
ACS sample.  Since field star contamination is critical in M55,
particularly in the external regions of the cluster, the definition of
the $B-V$ color edges of the selection box has required a detailed
study of the color-magnitude distribution of field stars.  To do this,
we have exploited both the outermost portion of the WFI observations
(beyond the tidal radius), and the Galaxy model of \citet{robin03} in
the direction of the cluster.  In order to limit both the risk of
field star and SGB blend contamination, we pick $B-V\simeq 0.41$ as a
conservative value for the red-edge of the BSS selection box.  As blue
limit, we have chosen $B-V\simeq 0.08$.  The adopted selection box in
the ($V,~B-V$) plane is shown in the right-hand panel of Figure
\ref{fig:optCMD}, and the number of enclosed BSS is 38.

Since M55 is known to harbor a large population of SX Phoenicis (SX Phe)
variables in the BSS region \citep{pych01}, we have cross-correlated the SX
Phe catalog with our data set.  All of the 24 SX Phe identified by
\citet{pych01} are contained in our sample (see {\it triangles} in
Figs. \ref{fig:uvCMD} and \ref{fig:optCMD}), and all but two lie within our
BSS selection boxes. The two outliers (V21 and V26 in their catalog) are
indeed only slightly redder than the adopted limits, and most likely are
genuine BSS \citep [in fact, BSS frequently show the pulsating properties of
SX Phe stars; see, e.g.,][]{mateo96}. Thus, they have been also included in
our BSS sample.
The SX~Phe population of M55 is truly remarkable, second only to the always
weird $\omega$ Centauri \citep{kalu04}. Considering that we have identified
56 BSS within the FoV in common with \citet{pych01} and that 24 of them are SX Phe
variables, we see that almost half ($43\%$) of the BSS in M55 are
pulsating.

The coordinates and magnitudes of all the selected BSS (65) are listed in
Table \ref{tab:BSS}, and their projected spatial distribution is shown in
Figures \ref{fig:HST} and \ref{fig:WFI}.  Two candidate BSS (namely, BSS 64
and 65 in Table \ref{tab:BSS}) lie at $r>r_t$. Since Z97 suggest that there
is tidal distortion in the north-east direction, these BSS could be part of a
cluster tidal tail. However, our observations do not indicate any significant
distortion in the cluster stellar distribution (although a more extended
mapping of the surrounding regions might be needed), and we therefore
conclude that they probably are field stars. Thus, they are not encircled in
the right-hand panel of Figure \ref{fig:optCMD}, and have not been considered
in the following analysis.

No quantitative comparison between our selected BSS population and that
presented in Z97 is possible, since they provide neither selection criteria
nor the coordinates of the identified BSS.  Within the FoV (the inner
$4\arcmin\times 4\arcmin$) in common with \citet{mandu97}, we find 33 BSS;
for comparison, by using the published BSS magnitudes, we have verified that
30 of their stars are included in our BSS selection boxes, thus showing a
very good agreement between the two studies.  The remaining 44 BSS identified
by these authors are fainter and/or redder than the limits adopted in the
present work.

\section{The Reference Populations}
To study the BSS projected radial distribution and detect possible peculiarities, a
reference population which is representative of the normal cluster stars must
be properly defined.  For this purpose we have chosen the horizontal branch
(HB) and the RGB populations, both of which are expected to have a
non-peculiar radial distribution within the cluster.

The adopted HB selection boxes in the optical CMDs are shown in Figure
\ref{fig:optCMD}, and are designed to include the bulk of this population.
The box in the UV plane defined by the stars in common between the ACS and
the WFPC2 samples is shown in Figure \ref{fig:uvCMD}, and confirms the
suitability of the adopted selection. By cross-correlating the coordinates of our 
catalog with
the catalog of RR Lyrae variables detected by \citet{olech99} we have
identified 10 stars ({\it filled squares} in Fig. \ref{fig:optCMD}) out of a
total of 13, the remaining three falling in the gaps of the WFI chips. The
few RR Lyrae that lie outside the selection boxes have been also included in
our HB sample. Thus, the total number of
selected HB stars is 237 (78 in the ACS data set, and 159 in the WFI sample)
\footnote{Only one object lying in the HB box has been excluded because
it is located at $r>r_t$}

In selecting the RGB sample, we have considered only the magnitude range
$15.8\le V\le 17.5$ (the same adopted for the BSS), since the brightest RGB
stars are saturated in the ACS observations. We have drawn narrow selection
boxes around the RGB mean ridge line in the CMDs, in order to limit the
contamination by field stars. The adopted boxes are shown in Figure
\ref{fig:optCMD}, and the resulting number of RGB stars found at $r\le r_t$
is 1504.

\subsection{Field Contamination}
\label{sec:deco}
As apparent from the right-hand panel of Figure \ref{fig:optCMD}, field star
contamination is a critical issue in the study of M55, particularly for the
cluster outer regions.  In order to estimate the impact of the field
contamination on the cluster population selections, we have considered the
CMD in the outermost ($r>r_t$) portion of the WFI data set. By considering
that the sampled area is of $\sim 252\,{\rm arcmin}^2$, counts of stars
within the adopted BSS, HB, and RGB selection boxes yield the number
densities of field stars contaminating the selected cluster populations. As a
further check, we have performed the same analysis on the synthetic data
Galaxy model of \citet{robin03} in the $B$ and $V$ bands, considering a much
larger area (1 square degree) in the direction of M55. The number densities
derived from the two methods agree within a factor of $\sim 2$--3, and we
have finally adopted densities obtained from the Galaxy model, because of the
much larger sampled area. The estimated contamination is roughly 8, 4, and
550 field stars per square degree for the selected populations of BSS, HB,
and RGB stars, respectively.  By using the $V$ and $I$ data of the synthetic
Galaxy model, we have verified that the same values are also appropriate for
decontaminating the cluster populations in the inner $202\arcsec\times
202\arcsec$ (the ACS FoV) of our sample, where the selection has been
performed in these photometric bands. The quoted values have been adopted in
the following analysis to statistically decontaminate the star counts.

\section{THE BSS PROJECTED RADIAL DISTRIBUTION}
\label{sec:radist}
As for other clusters studied in a similar way \citep[see references
in][]{lan07b}, we have searched for possible peculiarities in the BSS radial
distribution by comparing it with that of HB and RGB stars, that are expected
to be distributed as "normal" cluster stars.

We have used the Kolmogorov-Smirnov (KS) test to search for statistical
differences between the cumulative projected radial distributions of BSS and HB stars
(the comparison with the RGB population has not been performed because of the
non-negligible degree of field star contamination).  As shown in Figure
\ref{fig:KS}, BSS appear to be more concentrated than normal cluster stars
within $\sim 300\arcsec$ from the center, and less concentrated outward.  The
statistical significance of this result, however, is rather poor: the overall
KS probability that BSS and HB stars are not extracted from the same parent
population is $\sim 0.90$ (corresponding to $\sim 1.6\,\sigma$ significance
level). If the analysis is restricted to the inner $300\arcsec$, BSS are more
concentrated than HB stars at $\sim 1.9\,\sigma$ level. For $r>300\arcsec$,
where less than 20\% of the total BSS and HB populations are located, the BSS
are less concentrated at the $3\,\sigma$.  A similar trend, with a similar
statistical significance, was also found by Z97, who, however, performed the
comparison with the RGB population.

\begin{center}
\begin{figure}[!p]
\includegraphics[scale=0.7]{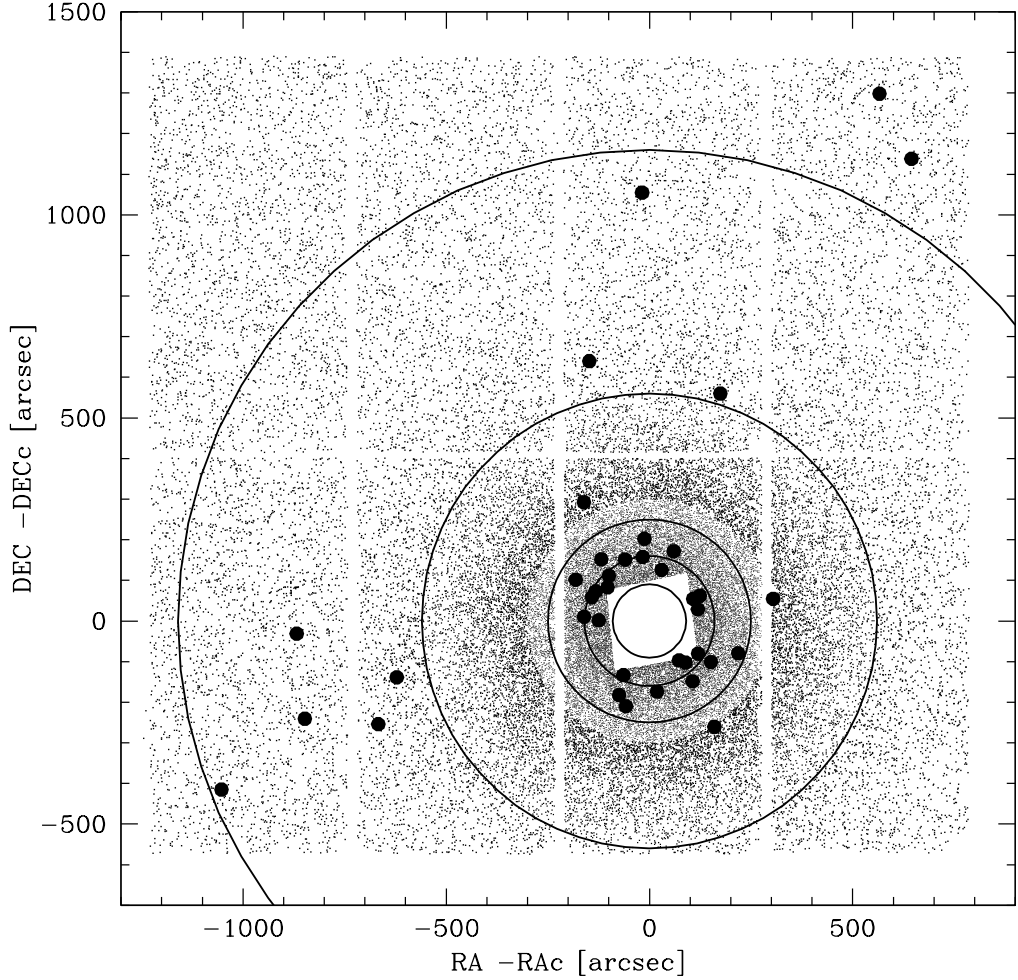}
\caption{Map of the complementary WFI sample, with coordinates referred to
$C_{\rm grav}$. The empty central region corresponds to the {\it HST}-ACS FoV
(dashed line in Fig. \ref{fig:HST}). All the detected BSS are marked as heavy
dots, and the concentric annuli used to study their projected radial distribution
(cfr. Table \ref{tab:annuli}) are shown as solid circles, with the inner
annulus corresponding to $r=90\arcsec$, and the outer one corresponding to
the tidal radius $r_t=1160\arcsec$. The two candidate BSS lying beyond $r_t$
most probably are field stars.}
\label{fig:WFI}
\end{figure}
\end{center}

\begin{center}
\begin{figure}[!p]
\includegraphics[scale=0.7]{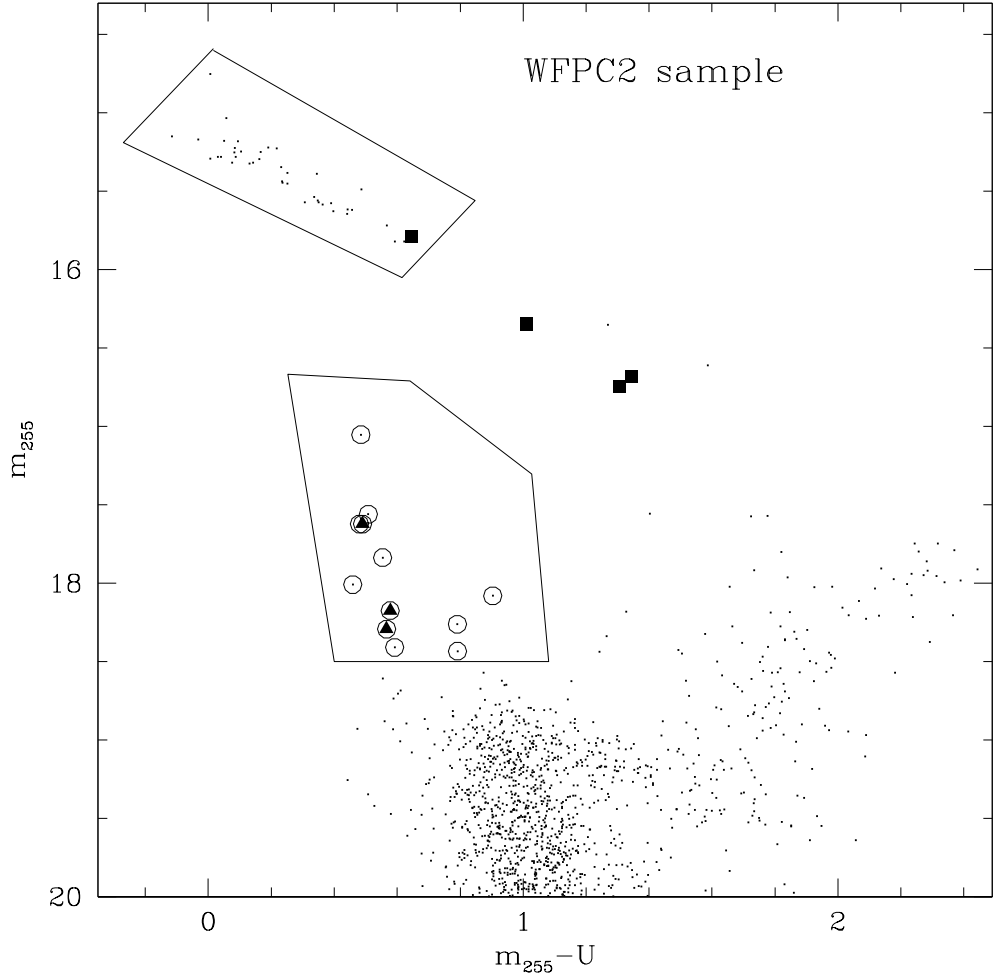}
\caption{Ultraviolet CMD of the {\it HST}-WFPC2 sample.  The adopted BSS and
HB selection boxes are shown.  The resulting fiducial BSS are marked with
{\it empty circles}. {\it Triangles} correspond to the SX Phoenicis variables
identified by \citet{pych01}, while the {\it squares} mark the RRLyrae
identified by \citet{olech99} and included in our HB sample.}
\label{fig:uvCMD}
\end{figure}
\end{center}

\begin{figure}[!p]
\begin{center}
\includegraphics[scale=0.7]{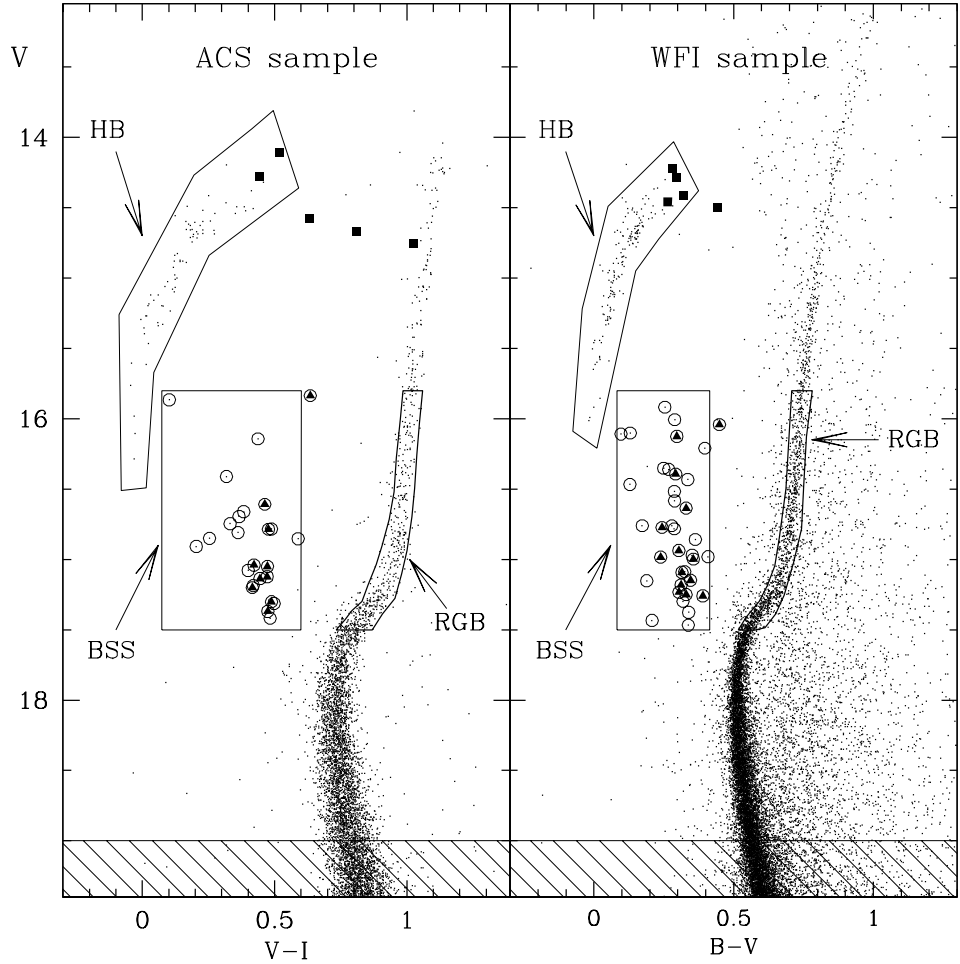}
\caption{Optical CMDs of the {\it HST}-ACS and of the complementary WFI
samples.  The adopted BSS, HB, and RGB selection boxes are shown. Symbols are
as in Figure \ref{fig:uvCMD}.  The hatched regions indicate the magnitude
limit ($V\le 19$) adopted for the computation of the cluster surface density
profile.}
\label{fig:optCMD}
\end{center}
\end{figure}

\begin{figure}[!p]
\begin{center}
\includegraphics[scale=0.7]{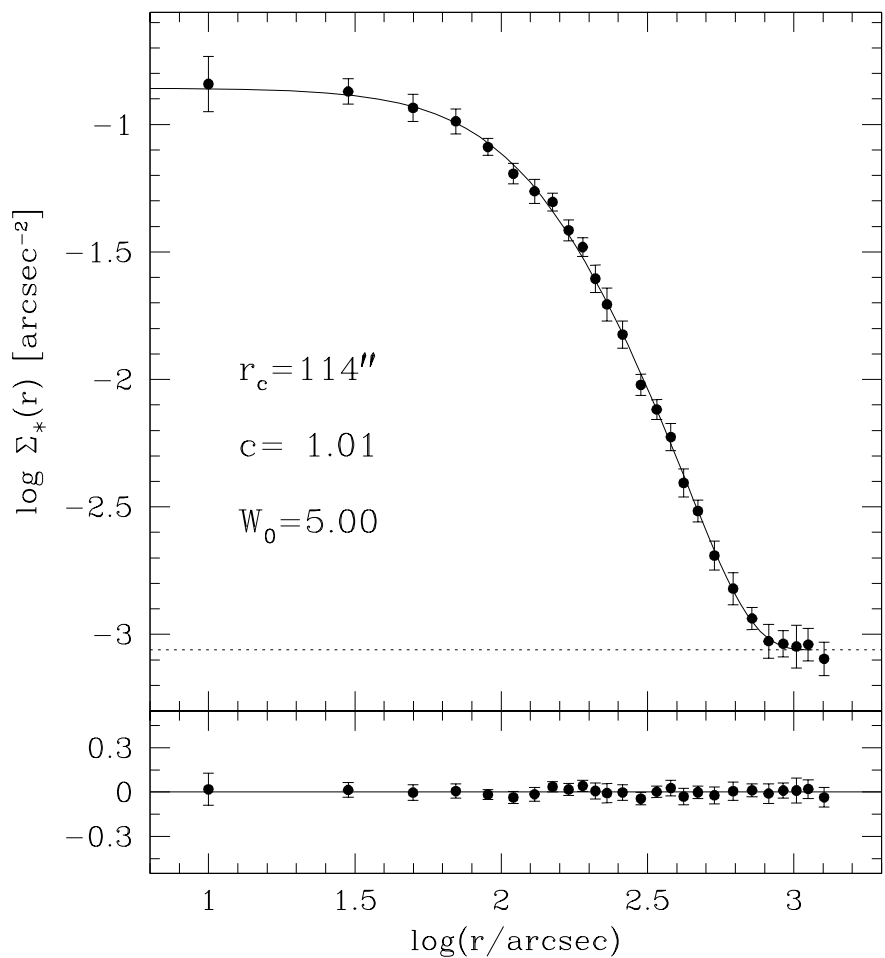}
\caption{Observed surface density profile ({\it dots and error bars}) and
best-fit King model ({\it solid line}). The radial profile is in units of
number of stars per square arcseconds.  The {\it dotted line} indicates the
adopted level of the background, and the model characteristic parameters
(core radius $r_c$, concentration $c$, dimensionless central potential $W_0$)
are marked in the figure. The lower panel shows the residuals between the
observations and the fitted profile at each radial coordinate.}
\label{fig:prof}
\end{center}
\end{figure}

\begin{figure}[!p]
\begin{center}
\includegraphics[scale=0.7]{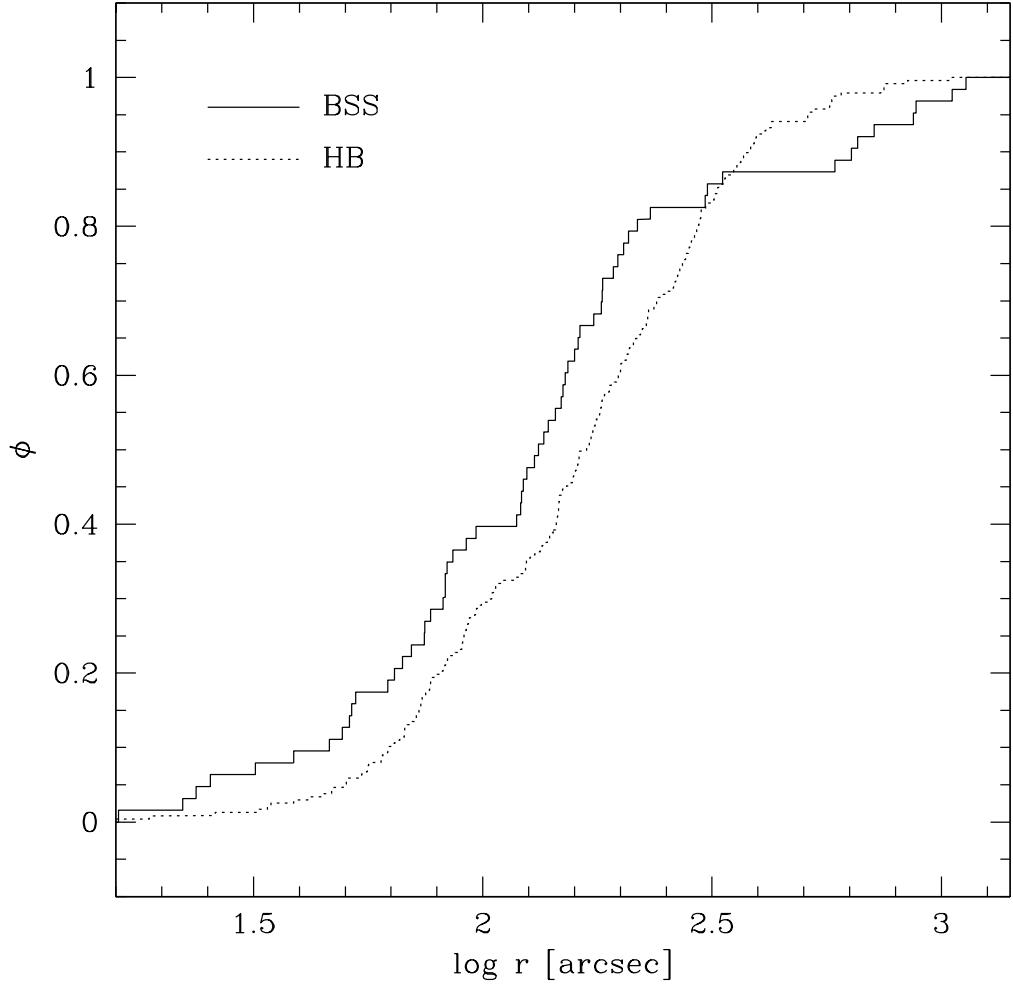}
\caption{Cumulative projected radial distribution of BSS ({\it solid line}) and HB
stars ({\it dotted line}).}
\label{fig:KS}
\end{center}
\end{figure}

For a more quantitative analysis, the surveyed area has been divided into 5
concentric annuli (see Figs. \ref{fig:HST} and \ref{fig:WFI}), and the number
of BSS, HB, and RGB stars ($N_{\rm BSS}$, $N_{\rm HB}$, and $N_{\rm RGB}$,
respectively) within each annulus has been counted. The resulting number
counts have then been corrected for field contamination by taking into
account the fraction of annulus area effectively sampled by our observations,
and the estimated density of contaminating field stars for each population
(see previous section). The values thus obtained are listed in Table
\ref{tab:annuli} and have been used to compute the specific frequencies
$N_{\rm BSS}/N_{\rm HB}$, $N_{\rm BSS}/N_{\rm RGB}$, and $N_{\rm HB}/N_{\rm
RGB}$. Since the number of stars in any post-MS stage is proportional to the
duration of the evolutionary phase itself \citep{renbuz86}, the specific
frequencies $N_{\rm HB}/N_{\rm RGB}$ is expected to be constant and equal to
the ratio between the evolutionary time scales of the HB phase and of the RGB
portion in the magnitude range $15.8\le V\le 17.5$, where the stars have been
counted.  In order to verify this, we have used the BASTi\footnote{Available
at {\tt http://www.te.astro.it/BASTI/index.php}} evolutionary model library
\citep[][and reference therein]{pietri06}, selecting the $\alpha-$enhanced
low-temperature opacities tracks computed for metallicities $\rm
[Fe/H]=-1.84$ and $\rm [M/H]=-1.49$ (the closest to the observed values $\rm
[Fe/H]=-1.61$ and $\rm [M/H]=-1.41$; Ferraro et al. 1999b).  From these models
we have estimated that the time spent by a 0.8\,$M_\odot$ star along the RGB
sequence in the range $15.8\le V\le 17.5$ is $t_{\rm RGB}\sim 0.6$ Gyr, while
the duration of the HB phase for a 0.63 $M_\odot$ is $t_{\rm HB}\sim 0.09$
Gyr; thus, $t_{\rm HB}/t_{\rm RGB}\simeq 0.15$, in good agreement with the
observed value of the $N_{\rm HB}/N_{\rm RGB}$ ratio (see the {\it dotted
line} in the lower panel of Figure \ref{fig:Npop}).  A very similar result is
also found by using the theoretical stellar tracks of the Pisa Evolutionary
Library\footnote{Available at {\tt http://astro.df.unipi.it/SAA/PEL/Z0.html}}
\citep[see references in] []{cariu04}, and it ensures that the selected (and
decontaminated) HB and RGB populations are indeed representative of the
normal cluster stars.  As for the BSS, the specific frequency $N_{\rm
BSS}/N_{\rm HB}$ shows a completely different projected radial distribution, with a
clearly bimodal behavior: from a central value of $\sim 0.4$, the BSS
specific frequency decreases to a minimum at about $4\, r_c$, and rises again
at larger radii. A very similar trend (with the central peak at $\sim 0.07$)
is also found for $N_{\rm BSS}/N_{\rm RGB}$, in agreement with Z97.

By integrating the density profile from the best-fit King model (see
Sect.\ref{sec:prof}), and assuming the values of central surface brightness,
reddening and distance modulus quoted in Sect.  \ref{sec:prof}, we have also
computed the luminosity sampled in each annulus ($L^{\rm samp}$), and the
total sampled luminosity ($L^{\rm samp}_{\rm tot}$) taking into account the
incomplete spatial coverage of the most external annulus (see
Fig. \ref{fig:WFI}).  The resulting ratios between these two quantities in
each annulus are listed in Table \ref{tab:annuli}, and have been used to
compute the double normalized ratio \citep[see][]{fe93}:
\begin{equation}
R_{\rm pop}=\frac{(N_{\rm pop}/N_{\rm pop}^{\rm tot})}{(L^{\rm samp}/L_{\rm
tot}^{\rm samp})},
\label{eq:spec_freq}
\end{equation}
where ${\rm pop} =$ BSS, HB, RGB.

The radial trend of $R_{\rm HB}$ (as well as that of $R_{\rm RGB}$) is
essentially constant, with a value close to unity (see Fig.
\ref{fig:Rpop}). This is just what expected on the basis of the stellar
evolution theory, which predicts that the fraction of stars in any post-MS
evolutionary stage is strictly proportional to the fraction of the sampled
luminosity \citep{renfusi88}.  Conversely,
the trend of $R_{\rm BSS}$ is bimodal and indicates that, with respect to the
sampled luminosity, the fraction of BSS is higher in the central regions and
(particularly) in the cluster outskirts, and smaller at intermediate radii,
with respect to the fraction of normal cluster stars

\section{DISCUSSION}
\label{sec:discussion}
We have found that the BSS projected radial distribution in M55 is bimodal, i.e.,
peaked in the center, decreasing at intermediate radii, and rising again in
the exterior. This is in agreement with the findings of Z97 from the analysis
of a much smaller fraction of the cluster, and puts their result on much more
solid statistical basis.

Such a bimodality is similar to that found in M3 \citep{fe97}, 47~Tuc
\citep{fe04}, NGC~6752 \citep{sab04}, and M5 \citep{w06,lan07a}. As in those
GCs, also in M55 the position of the observed minimum approximately
corresponds to the radius of avoidance $r_{\rm avoid}$ of the system, i.e.,
the radius within which all the stars as massive as $1.2\,M_\odot$ (which is
assumed to be the typical BSS mass) are expected to have already sunk to the
core due to dynamical friction and mass segregation processes.  In fact, by
using the dynamical friction timescale formula \citep[from, e.g.,][]{ma06}
with the best-fit King model and the central stellar density presented in
Sect. \ref{sec:prof}, and assuming $\sigma\simeq 4.9\,{\rm km\,s^{-1}}$ as
the central velocity dispersion \citep{pry93}, and 12 Gyr as the cluster age,
we estimate that $r_{\rm avoid}\simeq 4.5\, r_c$, in reasonable agreement
with the position of the observed minimum.

\begin{figure}[!p]
\begin{center}
\includegraphics[scale=0.7]{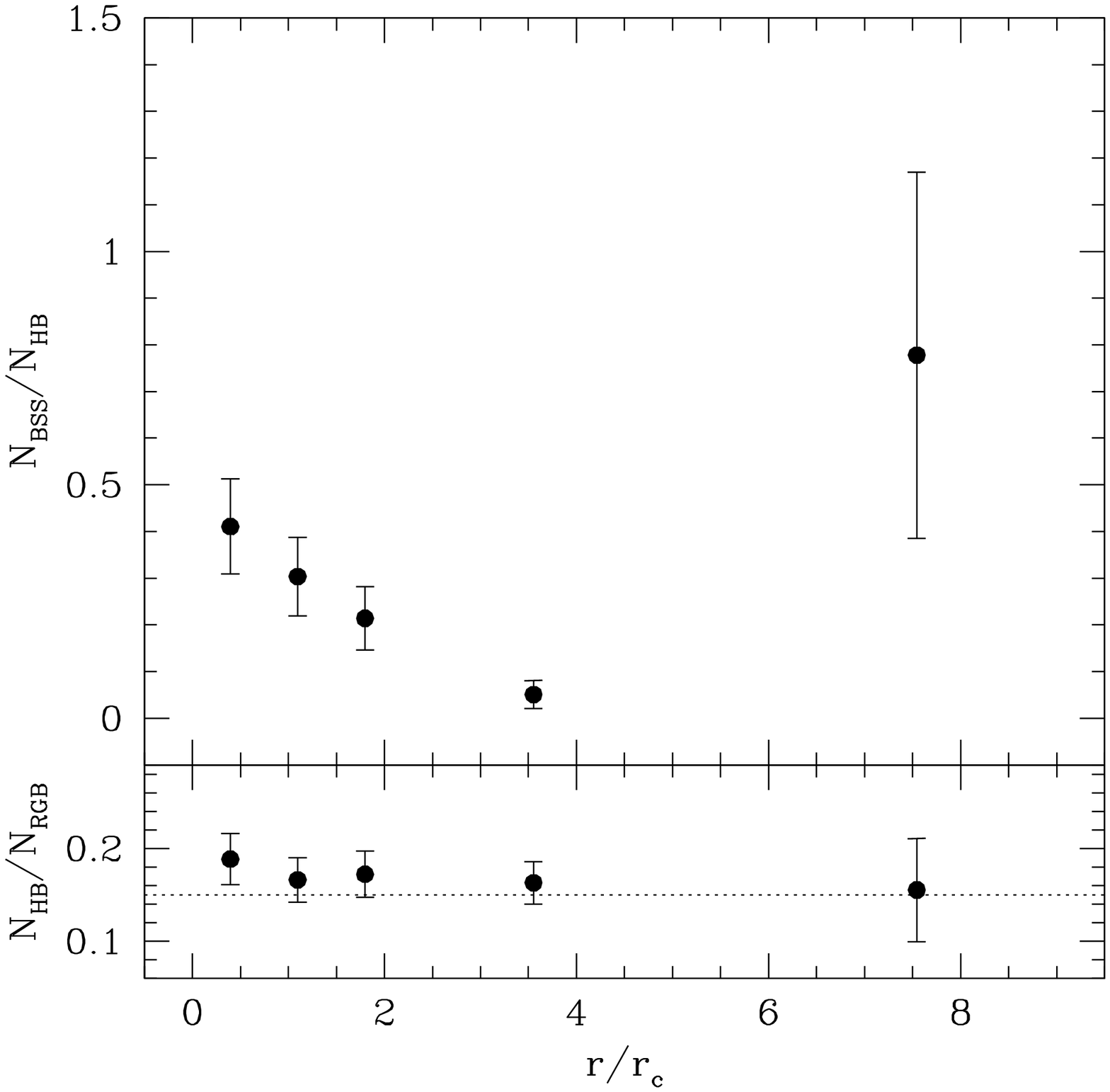}
\caption{{\it Upper panel:} Projected  radial distribution of the specific
frequency $N_{\rm BSS}/N_{\rm HB}$, as a function the radial distance from
the cluster center, expressed in units of the core radius.  {\it Lower
panel:} The same as above, for the specific frequency $N_{\rm HB}/N_{\rm
RGB}$. The {\it dotted line} corresponds to the value ($\sim 0.15$) predicted
by the population synthesis models of \citet{pietri06} for the ratio between
the evolutionary time-scales of the HB and RGB (in the range $15.8\le V\le
17.5$) phases.}
\label{fig:Npop}
\end{center}
\end{figure}

\begin{figure}[!p]
\begin{center}
\includegraphics[scale=0.7]{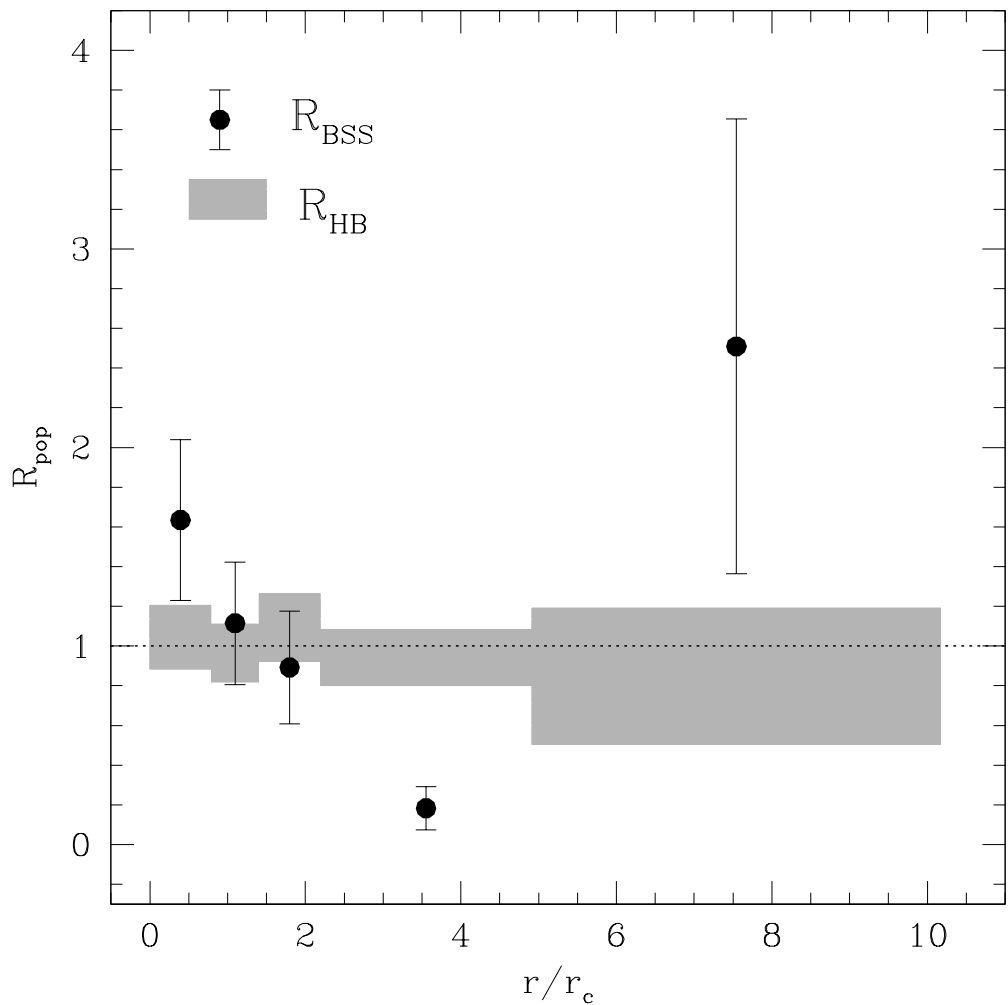}
\caption{Projected  radial distribution of the double normalized ratios of BSS
({\it dots}) and HB stars ({\it gray rectangles}), as defined in
equation~(\ref{eq:spec_freq}).  The error bars (represented by the vertical
sizes of the rectangles in the case of $R_{\rm HB}$) are computed as
described in \citet{sab04}.  The {\it dotted line} corresponds to the value
($R_{\rm pop}=1$) expected for any normal post-MS population in the cluster
(see Sect.\ref{sec:radist}).}
\label{fig:Rpop}
\end{center}
\end{figure}

The BSS specific frequency in the center of M55 ($N_{\rm BSS}/N_{\rm
HB}\simeq 0.4$) is also similar to that measured in the other bimodal GCs
\citep[cfr. Fig.\ref{fig:Npop}, with Fig. 12 of][and see also Lanzoni et
al. 2007b]{lan07a}, where the central peak of the distribution is found to be
mainly generated by COL-BSS \citep[see also][]{ma06}. However, the central
density in M55 is much lower (by a factor of 100 or more), and stellar
collisions are expected to be less important in this system. Indeed, the
cluster central density is quite similar to that of NGC~288 (only a factor of
two higher), where most of the central BSS are thought to be MT-BSS
\citep{bel02}. A remarkable difference in the central value of $N_{\rm
BSS}/N_{\rm HB}$ in these two low density clusters is however apparent. In
fact, by considering only the brightest portion of the BSS population in
NGC~288, \citet{fe03} measured $N_{\rm BSS}/N_{\rm HB}\simeq 1$, which is the
highest BSS frequency ever found in a GC, together with that of M80
\citep{fe99a}, and is more than twice that of M55.  What is the origin of
this difference?  One possibility is a different primordial binary
fraction. However, \citet{sollima07} have recently estimated that the binary
fractions in the core of the two clusters are the same ($\sim 10\%$).
Another possibility is a substantial difference in the collision rate.  By
using equation (14) from \citet{leon89}, we estimate that the central
binary-binary collision rate in M55 is only a factor of $\sim 2$ higher than
that in NGC~288. Moreover, the binary survival rate \citep[defined as the
ratio between the formation and destruction rates; see][]{verb03} is about
twice as high in M55 than in NGC~288.  Thus, our results indicate that two
clusters with similar environments (and collision rates) and similar
primordial binary content can produce quite different central BSS
populations.  Unfortunately, the BSS study in NGC~288 was restricted to two
WFPC2 frames, and an investigation covering the entire cluster extension is
urged in order to compare the global BSS population and its radial
distribution in the two systems.

Compared to the other bimodal GCs, the external rising branch in M55 is much
more prominent. It is the largest upturn found to date ($N_{\rm BSS}/N_{\rm
HB}\simeq 0.8\pm 0.4$ compared to the previous maximum value of $\simeq
0.25\pm 0.11$, found in 47~Tuc).  This is even more surprising if we consider
that only 10\% of the total cluster light is contained between $r_{\rm
avoid}$ and $r_t$ in M55, while it amounts to 32\% in the case of 47~Tuc.  As
discussed in \citet[][see also Lanzoni et al. 2007a]{ma06}, the external
rising branch is thought to be made of MT-BSS, generated in binary systems
evolving in isolation in the cluster outskirts \citep[this finding is also
confirmed by the recent N-body simulations of][]{hurley07}. Thus, such a
prominent upturn of the BSS distribution would imply a significantly higher
primordial binary fraction in M55, compared to the other GCs. This seems in
contrast with the results of \citet{sollima07}, who measured the binary
fractions in the core of 13 galactic GCs and found that M55 has one of the
lowest fractions ($\sim 10\%$), with respect to the others, which range up to
$\sim 50\%$ (in Terzan 7). However a better understanding of the evolution of
the binary fraction in the core, as a function of the cluster dynamical age,
is needed to better address this point. In fact, the theoretical expectations
for the time evolution of the core binary fraction are still controversial:
while \citet{ivanova05} suggest that such a fraction significantly decreases
in time, the opposite trend is found by \citet{hurley07}.  Moreover, since a
careful investigation of the BSS radial distribution has not yet been
performed in any of the other remaining clusters studied by
\citet{sollima07}, a comprehensive comparison of the BSS population
properties in these systems is not yet possible.

The nature of the central BSS and of those producing the external rising
branch in M55 is thus an open question.  Appropriate dynamical simulations
and detailed spectroscopic studies \citep[see, e.g.,][]{COdep} are therefore
urged.  We defer such studies to a forthcoming paper, where the results of
our entire sample of clusters will be compared and discussed.

\acknowledgements
This research was supported by Agenzia Spaziale Italiana
under contract ASI-INAF I/023/05/0, by the Istituto Nazionale di Astrofisica
under contract PRIN/INAF 2006, and by the Ministero dell'Istruzione,
dell'Universit\`a e della Ricerca. RTR is partially funded by NASA through
grant number GO-10524 from the Space Telescope Science Institute. ED is
supported by a grant financed by ASI.  We also acknowledge J. Kaluzny for
having provided us with the information about the FoV used in \citet{pych01}
and \citet{olech99}.  This research used the facilities of the Canadian
Astronomy Data Centre operated by the National Research Council of Canada
with the support of the Canadian Space Agency.

\clearpage

\begin{deluxetable}{lcccccccc}
\tabletypesize{\footnotesize}
\tablewidth{15.5cm} 
\tablecaption{The BSS Population of M55}
\tablehead{
\colhead{Name} &
\colhead{RA} &
\colhead{DEC} &
\colhead{$m_{255}$} &
\colhead{U} &
\colhead{B} &
\colhead{V} &
\colhead{I} &
\colhead{SX Phe}\\
\colhead{ } &
\colhead{[degree]} &
\colhead{[degree]} &
\colhead{ } &
\colhead{ } &
\colhead{ } &
\colhead{ } &
\colhead{ } &
\colhead{ }
}
\startdata 
BSS  1 & 294.9998920 & -30.9667245 & 18.26 & 17.47 &   -   &   -   &   -   &  -  \\ 
BSS  2 & 294.9954953 & -30.9396261 & 18.08 & 17.18 &   -   & 16.85 & 16.26 &  -  \\ 
BSS  3 & 295.0121689 & -30.9581611 & 17.05 & 16.57 &   -   & 16.14 & 15.71 &  -  \\ 
BSS  4 & 294.9982015 & -30.9483228 & 17.56 & 17.05 &   -   & 16.66 & 16.28 &  -  \\ 
BSS  5 & 295.0166912 & -30.9704646 & 17.62 & 17.14 &   -   & 16.69 & 16.33 &  -  \\ 
BSS  6 & 295.0193344 & -30.9660591 & 17.84 & 17.28 &   -   & 16.81 & 16.45 &  -  \\ 
BSS  7 & 295.0045478 & -30.9669382 & 18.01 & 17.55 &   -   & 16.91 & 16.70 &  -  \\ 
BSS  8 & 295.0033265 & -30.9834341 & 18.43 & 17.64 &   -   & 17.19 & 16.77 &  -  \\ 
BSS  9 & 295.0104327 & -30.9803687 & 18.41 & 17.82 &   -   & 17.42 & 16.93 &  -  \\ 
BSS 10 & 295.0122305 & -30.9747912 & 17.62 & 17.13 &   -   & 16.61 & 16.15 & V41 \\ 
BSS 11 & 295.0040550 & -30.9659563 & 18.29 & 17.73 &   -   & 17.30 & 16.81 & V31 \\ 
BSS 12 & 294.9902115 & -30.9506018 & 18.17 & 17.60 &   -   & 17.37 & 16.90 & V19 \\ 
BSS 13 & 294.9849393 & -30.9719486 &   -   &   -   &   -   & 15.87 & 15.76 &  -  \\ 
BSS 14 & 294.9748793 & -30.9741471 &   -   &   -   &   -   & 16.41 & 16.09 &  -  \\ 
BSS 15 & 294.9858796 & -30.9600404 &   -   &   -   &   -   & 16.85 & 16.60 &  -  \\ 
BSS 16 & 294.9702800 & -30.9607749 &   -   &   -   &   -   & 17.08 & 16.68 &  -  \\ 
BSS 17 & 295.0254214 & -30.9727945 &   -   &   -   &   -   & 16.78 & 16.30 &  -  \\ 
BSS 18 & 295.0100545 & -30.9422429 &   -   &   -   &   -   & 16.75 & 16.41 &  -  \\ 
BSS 19 & 295.0214094 & -30.9804905 &   -   &   -   &   -   & 17.31 & 16.82 &  -  \\ 
BSS 20 & 294.9951572 & -30.9710352 &   -   &   -   &   -   & 16.78 & 16.31 & V38 \\ 
BSS 21 & 294.9921499 & -30.9759958 &   -   &   -   &   -   & 17.04 & 16.62 & V32 \\ 
BSS 22 & 295.0285621 & -30.9424951 &   -   &   -   &   -   & 17.05 & 16.58 & V18 \\ 
BSS 23 & 294.9788871 & -30.9728224 &   -   &   -   &   -   & 17.12 & 16.65 & V20 \\ 
BSS 24 & 294.9751258 & -30.9689860 &   -   &   -   &   -   & 17.14 & 16.69 & V27 \\ 
BSS 25 & 294.9941390 & -30.9568945 &   -   &   -   &   -   & 17.20 & 16.78 & V42 \\ 
BSS 26 & 294.9927055 & -30.9852788 &   -   &   -   &   -   & 15.84 & 15.20 & V21 \\ 
BSS 27 & 294.9793701 & -31.0208092 &   -   &   -   & 16.17 & 15.92 &   -   &  -  \\ 
BSS 28 & 294.7966919 & -31.0010357 &   -   &   -   & 16.21 & 16.11 &   -   &  -  \\ 
BSS 29 & 295.0544434 & -30.8069954 &   -   &   -   & 16.23 & 16.10 &   -   &  -  \\ 
BSS 30 & 295.0268860 & -30.9911098 &   -   &   -   & 16.30 & 16.01 &   -   &  -  \\ 
BSS 31 & 295.0368652 & -30.9847641 &   -   &   -   & 16.60 & 16.47 &   -   &  -  \\ 
BSS 32 & 295.0367126 & -30.9545650 &   -   &   -   & 16.60 & 16.35 &   -   &  -  \\ 
BSS 33 & 295.0966492 & -30.9473000 &   -   &   -   & 16.61 & 16.21 &   -   &  -  \\ 
BSS 34 & 294.9940796 & -30.9063625 &   -   &   -   & 16.63 & 16.36 &   -   &  -  \\ 
BSS 35 & 294.9552917 & -30.9421539 &   -   &   -   & 16.77 & 16.43 &   -   &  -  \\ 
BSS 36 & 295.0687561 & -30.9846306 &   -   &   -   & 16.81 & 16.52 &   -   &  -  \\ 
BSS 37 & 295.0217285 & -30.9895248 &   -   &   -   & 16.87 & 16.58 &   -   &  -  \\ 
BSS 38 & 294.7169495 & -30.9712677 &   -   &   -   & 16.93 & 16.76 &   -   &  -  \\ 
BSS 39 & 294.9458923 & -30.8814220 &   -   &   -   & 17.04 & 16.76 &   -   &  -  \\ 
BSS 40 & 294.9922485 & -30.6695671 &   -   &   -   & 17.07 & 16.78 &   -   &  -  \\ 
BSS 41 & 295.0174561 & -30.9149532 &   -   &   -   & 17.22 & 16.86 &   -   &  -  \\ 
BSS 42 & 294.7818298 & -31.0331841 &   -   &   -   & 17.32 & 16.97 &   -   &  -  \\ 
BSS 43 & 294.7232361 & -31.0292740 &   -   &   -   & 17.34 & 17.15 &   -   &  -  \\ 
BSS 44 & 294.9502563 & -30.7848854 &   -   &   -   & 17.39 & 16.98 &   -   &  -  \\ 
BSS 45 & 294.9739380 & -31.0131721 &   -   &   -   & 17.41 & 17.09 &   -   &  -  \\ 
BSS 46 & 295.0329895 & -30.9473553 &   -   &   -   & 17.62 & 17.30 &   -   &  -  \\ 
BSS 47 & 294.6565247 & -31.0778027 &   -   &   -   & 17.64 & 17.43 &   -   &  -  \\ 
BSS 48 & 294.9787292 & -30.9204979 &   -   &   -   & 17.72 & 17.38 &   -   &  -  \\ 
BSS 49 & 294.9926758 & -30.9187489 &   -   &   -   & 17.81 & 17.47 &   -   &  -  \\ 
BSS 50 & 294.9646912 & -30.9394836 &   -   &   -   & 16.42 & 16.13 &   -   & V25 \\ 
BSS 51 & 294.9772644 & -30.9996738 &   -   &   -   & 16.69 & 16.39 &   -   & V33 \\ 
BSS 52 & 294.9597473 & -30.9203262 &   -   &   -   & 16.97 & 16.64 &   -   & V35 \\ 
BSS 53 & 294.9522705 & -30.9460793 &   -   &   -   & 17.02 & 16.77 &   -   & V36 \\ 
BSS 54 & 295.0324402 & -31.0037651 &   -   &   -   & 17.24 & 16.94 &   -   & V22 \\ 
BSS 55 & 294.9576721 & -30.9620571 &   -   &   -   & 17.22 & 16.98 &   -   & V37 \\ 
BSS 56 & 295.0382690 & -30.9452572 &   -   &   -   & 17.35 & 17.00 &   -   & V16 \\ 
BSS 57 & 294.9394226 & -30.9343033 &   -   &   -   & 17.41 & 17.09 &   -   & V24 \\ 
BSS 58 & 295.0471497 & -30.9905624 &   -   &   -   & 17.49 & 17.15 &   -   & V17 \\ 
BSS 59 & 295.0498962 & -31.0348148 &   -   &   -   & 17.49 & 17.18 &   -   & V39 \\ 
BSS 60 & 295.0078125 & -30.9275074 &   -   &   -   & 17.54 & 17.23 &   -   & V40 \\ 
BSS 61 & 295.0041809 & -31.0107975 &   -   &   -   & 17.58 & 17.25 &   -   & V34 \\ 
BSS 62 & 294.9658203 & -30.9315720 &   -   &   -   & 17.65 & 17.26 &   -   & V23 \\ 
BSS 63 & 294.9459534 & -30.9596825 &   -   &   -   & 16.49 & 16.04 &   -   & V26 \\ 
BSS 64 & 295.2062378 & -30.6464367 &   -   &   -   & 17.82 & 17.41 &   -   &  -  \\ 
BSS 65 & 295.1806946 & -30.6018009 &   -   &   -   & 16.53 & 16.34 &   -   &  -  
\enddata
\tablecomments{The first 12 BSS have been identified in the WFPC2; BSS 2--26
are from the ACS observations, the first 11 being in common with the WFPC2
sample; BSS 27--65 are from the complementary WFI data set.  BSS 64 and 65
lie beyond the cluster tidal radius, at $\sim 22\arcmin$ and $24\arcmin$ from
the center, respectively, and have not been considered in the analysis of the
BSS radial distribution.  The last column list the corresponding SX Phe stars
identified by \citet{pych01}.}
\label{tab:BSS}
\end{deluxetable}

\begin{deluxetable}{rrrrrrrc}
\tablecaption{Number Counts of BSS, HB, and RGB Stars, and Fraction of
Sampled Luminosity}
\tabletypesize{\scriptsize}
\tablewidth{0cm}
\tablehead{
\colhead{$r_i\arcsec$} &
\colhead{$r_e\arcsec$} &
\colhead{ } &
\colhead{$N_{\rm BSS}$} &
\colhead{$N_{\rm HB}$} &
\colhead{ } &
\colhead{$N_{\rm RGB}$} &
\colhead{$L^{\rm samp}/L_{\rm tot}^{\rm samp}$}
}
\startdata
   0 &   90 && 23 ~~~   &  56 ~~~    && 297~$\,$(1) & 0.23 \\
  90 &  160 && 17 ~~~   &  56 ~~~    && 337~$\,$(2) & 0.25 \\
 160 &  250 && 12 ~~~   &  56 ~~~    && 325~$\,$(5) & 0.22 \\
 250 &  560 &&  3 ~~~$\,$& 59 ~~~    && 362$\,$(33) & 0.26 \\
 560 & 1160 &&  7$\,$(1)&   9$\,$(1) &&  58$\,$(84) & 0.04 \\
\hline 
\enddata
\tablecomments{The values listed out of the parenthesis correspond to the
number of stars assumed to belong to the cluster (and thus used in the analysis),
while those in the parenthesis are estimated to be contaminating field
stars (see Sect. \ref{sec:deco}).}
\label{tab:annuli}
\end{deluxetable}

\clearpage

\end{document}